
  \parindent=0 true cm


  \def\sa{\vskip 0.30 true cm}
  \def\sb{\vskip 0.60 true cm}

  \baselineskip=0.60 true cm

\rightline{\bf LYCEN 9002}

\rightline{(January 1990)}

\sa
\sb
\sa

\centerline{\bf PERIODICITY AND QUASI-PERIODICITY}

\centerline{\bf FOR SUPER-INTEGRABLE HAMILTONIAN SYSTEMS}

\sa
\sb

\centerline{{\bf M. KIBLER}\footnote*{Permanent address : 
Institut de Physique Nucl\'eaire de 
Lyon, IN2P3-CNRS et Universit\'e Claude Bernard, F-69622 
Villeurbanne Cedex, France.} and {\bf P. WINTERNITZ}}

\sa

\centerline{Centre de recherches math\'ematiques,} 

\centerline{Universit\'e de Montr\'eal,}

\centerline{C.P. 6128-A, Montr\'eal, Qu\'ebec, Canada H3C 3J7} 

\sa
\sb
\sb

\parindent = 1 true cm
\baselineskip = 0.65 true cm

\centerline{ABSTRACT}

\sa

Classical trajectories are calculated for two Hamiltonian 
systems with ring shaped potentials. Both systems are 
super-integrable, but not maximally super-integrable, having 
four globally defined single-valued integrals of motion each. 
All finite trajectories are quasi-periodical; they become 
truly periodical if a commensurability condition is imposed on 
an angular momentum component.

\sa
\sb

\centerline{R\'ESUM\'E}

\sa

Les trajectoires classiques sont calcul\'ees pour deux 
syst\`emes hamiltoniens avec des potentiels en forme d'anneau. 
Les deux syst\`emes consid\'er\'es sont super-int\'egrables, 
mais pas de fa\c con maximale, ayant chacun quatre int\'egrales 
de mouvement uni-valu\'ees et d\'efinies globalement. Toutes 
les trajectoires finies sont quasi-p\'eriodiques. Elles deviennent 
p\'eriodiques si l'on impose une condition de 
commensurabilit\'e sur l'une des composantes du moment 
angulaire.

\sa
\sb
\sb

(published in Phys. Lett. {\bf A147}, 338-342 (1990))

\vfill\eject

\parindent = 1 true cm
  \baselineskip = 0.83 true cm

\noindent{\bf 1. Introduction}

\sa

The purpose of this letter is to discuss classical solutions 
for two super-integrable Hamiltonian systems with ring shaped potentials. 
The relevant Hamiltonian (i.e., the Hamilton function in 
classical mechanics or the usual Hamiltonian in 
non-relativistic quantum mechanics) can in both cases be written as 
$$
H = {1 \over 2} {\vec p}\,^2 + V({\vec r}), 
\eqno (1)
$$
where the two potentials are 
$$
V_O = {1\over 2} \Omega^2 (x^2+y^2+z^2) + {1\over 2} Q {1\over {x^2+y^2}}, 
\qquad \Omega > 0, \qquad Q \geq 0 
\eqno (2)
$$
and 
$$
V_C = -Z {1\over {\sqrt{x^2+y^2+z^2}}} + {1\over 2} Q 
{1\over {x^2+y^2}}, \qquad Z > 0, \qquad Q \geq 0.
\eqno (3)
$$
(The reduced mass of the particle moving in $V_O$ or $V_C$ is 
taken to be equal to 1.) 
The potentials $V_O$ and $V_C$ are cylindrically symmetrical, 
with $O(2)$ as geometrical symmetry group. 
In the limiting case $Q = 0$, $V_O$ goes over into the potential of 
an isotropic harmonic oscillator, whereas $V_C$ reduces to an 
attractive Coulomb potential.

\sa

The two considered systems thus extend the only two 
three-dimensional systems with the following properties.

\noindent (i) All finite classical trajectories are closed [1].

\noindent (ii) The systems are ``maximally super-integrable'' [2]. This 
means that each of the associated classical systems has $2n - 1 
= 5$ functionally independent single-valued integrals of motion, 
globally defined on the phase space in $2n = 6$ dimensions ($n$ 
is the number of degrees of freedom). Moreover, from these 
integrals of motion, we can construct in each case at least two 
different sets of $n = 3$ integrals of motion in involution, 
having only the Hamilton function in common. Indeed, two such 
sets are, for instance, 
$$
\{H; \ X_1 = {1\over 2} p_x^2 + {1\over 2} \Omega^2 x^2; 
     \ X_2 = {1\over 2} p_y^2 + {1\over 2} \Omega^2 y^2\}, 
\qquad \{H; \ {\vec L}^2; \ L_z\}
\eqno (4)
$$
for the harmonic oscillator system and 
$$
\{H; \ {\vec L}^2; \ L_x^2 + \tau L_y^2, 
\ 0 < \tau < 1\}, \qquad \{H; \ L_z; \ L_xp_y - L_yp_x + 
Z {z \over {\sqrt{x^2+y^2+z^2}}}\}
\eqno (5)
$$
for the Coulomb system. (We use $p_i$ and $L_i$ to denote linear 
and angular momenta, respectively. All constants of motion are 
given in classical form ; the corresponding quantum mechanical 
form follow from ordinary symmetrization.)

\noindent (iii) The energy levels for each of the associated quantum 
mechanical systems are degenerate. The level degeneracy is 
described for both systems by irreducible representation 
classes of a dynamical symmetry group with Lie algebra of rank 
two : $O(4)$ for the hydrogen atom [3,4,5] and $SU(3)$ for the 
harmonic oscillator [6,7].

\sa

A systematic search for super-integrable 
Hamiltonian systems in two [8] and 
three [9] dimensions was initiated some time ago. The 
corresponding integrals of motion were restricted to being 
second-order polynomials in the momenta and this related the 
problem to that of separating variables [8,9,10] in 
Hamilton-Jacobi or Schr\"odinger equations. In the 
three-dimensional case, it was shown that systems that have two 
different pairs of commuting integrals of motion quadratic in 
the momenta allow the separation of variables in two systems of 
coordinates. The requirement that the four integrals involved 
be functionally independent was not imposed. Hence, while all 
systems obtained in ref.~[9] were super-integrable (having more 
than $n = 3$ integrals of motion), they were not necessarily 
maximally super-integrable (having generally less than $2n - 1 = 5$ 
integrals of motion).

\sa

The potentials $V_O$ and $V_C$ (see (2) and (3)) correspond to 
such super-integrable systems, each having four (rather than 
five, the maximal number) functionally independent 
integrals of motion that happen to be second-order polynomials 
in the momenta. Both corresponding dynamical systems are of 
physical interest and have been studied before, mainly in their 
quantum mechanical incarnation. Thus, the potential $V_C$ is the 
so-called Hartmann potential, introduced in quantum chemistry 
for describing ring shaped molecules like cyclic polyenes [11]. 
The degeneracy of 
its quantum energy levels has been described in terms of an 
underlying $SU(2)$ dynamical symmetry group [12,13]. The 
potential $V_O$ might be useful in a nuclear physics context. 
Its levels also manifest an $SU(2)$ degeneracy [14]. It is to 
be noted that two systems that 
resemble the $V_C$ and the $V_O$ 
systems, viz., the $ABC$ [15] and the $ABO$ 
[16] systems, have been recently investigated in connection with 
the Aharonov-Bohm effect.

\sb

\noindent{\bf 2. Classical trajectories} 

\sa

\noindent{2.a The potential $V_O$}

\sa

The potential (2) is a special case of the potential 
$$
V_3 = \alpha (x^2+y^2+z^2) + \beta {1\over {z^2}} 
                           + h({y\over x}) {1\over {x^2+y^2}} 
\eqno (6)
$$
introduced by Makarov {\it et al.} [9]. In the most general 
case where $h$ is an arbitrary function, this potential allows 
the separation of variables for the Hamilton-Jacobi and the 
Schr\"odinger equations in four systems of coordinates, namely 
spherical, cylindrical, prolate spheroidal and oblate 
spheroidal coordinates ; for $h = 0$, we also have separability in 
Cartesian coordinates. The system corresponding to the 
Hamiltonian (1) with potential (2) admits the following four 
functionally independent integrals of motion
$$
H, \qquad 
A_1 = L_x^2 + L_y^2 + L_z^2 + Q(1 + {z^2 \over {x^2+y^2}}), \qquad
A_2 = {1\over 2} (p_z^2 + \Omega^2 z^2), \qquad 
A_3 = L_z.
\eqno (7)
$$
Separation of variables in the four systems mentioned above 
corresponds to the use of the four following sets of integrals 
of motion in involution
$$
\{H ; \ A_1 ; \ A_3\}, \qquad 
\{H ; \ A_2 ; \ A_3\}, \qquad 
\{H ; \ A_1 \mp 2a^2(H-A_2), \ a \in {\bf R} ; \ A_3\},
\eqno (8)
$$
respectively. 

\sa

In order to obtain the classical trajectories, we make use of 
the triplet $\{H; A_2; A_3\}$ corresponding to circular cylindrical 
coordinates ($x = \rho \cos \varphi, \ y = \rho \sin \varphi, \ 
z$). We write the integrals as 
$$H - A_2 = {1\over 2} ({\dot \rho}^2 + {{m^2 + Q}\over \rho^2} 
+ \Omega^2 \rho^2) \equiv E_1,$$
$$A_2 = {1\over 2} ({\dot z}^2 + \Omega^2 z^2) \equiv E_2, 
\eqno (9)$$
$$A_3 = \rho^2 {\dot \varphi} \equiv m.$$ 
We see that the trajectories are always finite, satisfying
$$\rho_1 \le \rho \le \rho_2, \qquad -z_0 \le z \le z_0,$$
$$(\rho_{1,2})^2 = {1 \over \Omega^2} 
\left( E_1 \mp \sqrt{E_1^2 - \Omega^2 M^2} \right), 
\qquad z_0 = \sqrt{2} {\sqrt{E_2} \over \Omega}, \eqno (10)$$
$$E_1 \ge \Omega \vert M \vert, \qquad 
\vert M \vert = \sqrt{m^2 + Q}, \qquad E_2 \ge 0.$$
Solving the first order ordinary differential equations (9), we 
obtain the general expression for the trajectories :
$$\rho(t) = {1 \over \sqrt{2}} \sqrt {\rho_1^2 + 
                                  \rho_2^2 +
                                 (\rho_2^2 - 
                                  \rho_1^2) 
\sin [2 \Omega(t-t_0)]},$$
$$z(t) = z_0 \sin [\Omega(t-t_0^{\prime})], \eqno (11)$$
$$\varphi(t) = 
  \varphi_0 + {1 \over 2} {m \over {\vert M \vert}} 
\sin^{-1} \left\{ 
{{(\rho_2^2 + \rho_1^2) \sin [2 \Omega(t-t_0)] - \rho_1^2 + \rho_2^2} \over 
 {(\rho_2^2 - \rho_1^2) \sin [2 \Omega(t-t_0)] + \rho_1^2 + \rho_2^2}} 
\right\},$$
where $t_0$, $t_0^{\prime}$ and $\varphi_0$ are further integration 
constants (in addition to the angular momentum component $m$ 
and the oscillation energies $E_1$ and $E_2$).

\sa

For $Q=0$, all trajectories are planar and closed (ellipses) as 
expected for the harmonic oscillator system ; 
the motion has the period $T_O = 2 \pi / \Omega$. 
Interestingly enough, for strictly positive values of $Q$ this 
is no longer the case. The coordinates $\rho$ and $z$ are 
``libration'' coordinates [17], for which periodicity implies 
$\rho(t + T_{\rho}) = \rho(t)$ and $z(t + T_z) = z(t)$. In the case 
under consideration we have $T_z = 2 T_\rho = T_O$, in function of the 
harmonic oscillator period $T_O$. The coordinate $\varphi$ is an angular 
one, so periodicity means 
$\varphi(t + T_\varphi) = \varphi(t) \pm 2 \pi$. It is clear 
that for arbitrary values of $m$, the durations $T_z$, 
$T_{\rho}$ 
and $T_\varphi$ are not necessarily commensurable. All trajectories are 
thus quasi-periodic : periodic in each coordinate, but not 
periodic in a global way. The condition for genuine periodicity 
is 
$$
{{\vert M \vert} \over m} = {k_1 \over k_2} 
\quad \Rightarrow \quad 
m^2 = {k_2^2 \over {k_1^2 - k_2^2}} Q,
\eqno (12)
$$
where $k_1$ and $k_2$ are mutually prime integers (with 
$\vert k_1 / k_2 \vert \ge 1$). The overall period $T$
then is
$$
T = 2 k_1 T_{\rho} = k_1 T_O.
\eqno (13)
$$
Therefore, for fixed $Q > 0$, in order for the trajectories to 
be periodic, rather than quasi-periodic, we obtain a ``classical 
quantization'' condition (see eq.~(12)) for the angular 
momentum component $L_z$.

\sa

In general the periodic and quasi-periodic 
trajectories do not lie in a plane (except for 
$Q= 0$). If $Q > 0$, there are at least two cases for which the 
trajectories are planar.

\noindent (i) For $m = 0$ : then from (11) we get $\varphi = \varphi_0$ 
and we see that in addition to being planar the motion is periodic of 
period $T_O$.

\noindent (ii) For $E_2 = 0$ : then from (10) and (11), the 
motion is in the $xy$-plane and is periodic if and only if 
eq.~(12) is satisfied.

\noindent The planarity of the trajectories for the potential 
$V_O$ (and for $V_C$ too) will be studied in a systematic way by calculating 
the torsion of the relevant curves [18]. 

\sb

\noindent{2.b The potential $V_C$}

\sa

The potential (3) is a special case of the potential 
$$
V_4 = \alpha {1\over r} 
      + \beta {\cos \theta \over {r^2 \sin^2 \theta}} 
      + h(\varphi)       {1\over {r^2 \sin^2 \theta}}
\eqno (14)
$$
found by Makarov {\it et al.} [9], allowing the separation of 
variables in spherical and parabolic coordinates. The 
Hamiltonian (1) with potential $V_C$ permits four functionally 
independent globally defined single-valued integrals of motion, 
namely
$$
H, \quad 
B_1 = L_x^2 + L_y^2 + L_z^2 + Q {1 \over {\sin^2{\theta}}}, \quad 
B_2 = L_z, \quad B_3 = L_xp_y - L_yp_x + Z {z \over r} 
- Q {z \over {x^2 + y^2}}.
\eqno (15)
$$
Separation of variables 
in spherical or parabolic coordinates 
corresponds to the simultaneous diagonalization of 
$$
\{ H ; B_1 ; B_2 \} \quad {\rm or} \quad \{ H ; B_2 ; B_3 \}, 
\eqno (16)
$$
respectively. 

\sa

In a previous article [13] we made use of the 
integrals $B_2$ and $B_3$ to obtain classical trajectories in 
parabolic coordinates. Here, we switch to spherical coordinates 
($x= r \sin \theta \cos \varphi, \ y = r \sin \theta \sin 
\varphi, \ z = r \cos \theta$) and use the triplet 
$\{H ; B_1 ; B_2\}$ to put
$$H = {1 \over 2} {\dot r}^2 + {1 \over 2} B_1 {1 \over r^2} - 
Z {1 \over r} \equiv E,$$
$$B_1 = r^4 \, {\dot \theta}^2 + {M^2 \over {\sin^2 \theta}} 
\equiv K, \eqno (17)$$
$$B_2 = r^2 \> {\sin^2 \theta} \> {\dot \varphi} \equiv m,$$
where again $M^2 = m^2 + Q$. From 
the equation for \ $\dot r$ \ in (17) we see that the 
motion is infinite for $E \ge 0$. Moreover, the finite 
trajectories correspond to
$$r_1 \le r \le r_2, \qquad 
\theta_0 \le \theta \le \pi - \theta_0,$$
$$r_{1,2} = {1 \over {-2E}} \left( Z \mp \sqrt{Z^2 + 2 E K} \right), 
\qquad 
\sin \theta_0 = {{\vert M \vert} \over {\sqrt{K}}}, \eqno (18)$$
$$- {Z^2 \over {2 K}} \le E < 0, \qquad K \ge M^2.$$
Equations (17) can be integrated to give
$$t-t_0 = - (-2E)^{-1/2} \sqrt{(r - r_1)(r_2 - r)} + 
Z (-2E)^{-3/2} \sin^{-1} \left( {2 r \over {r_2 - r_1}} 
- {{r_2 + r_1} \over {r_2 - r_1}} \right),$$
$$r \cos \theta = \cos {\theta_0} {1 \over {r_2 - r_1}}
\left\{ [2 r_1 r_2 - (r_1 + r_2)r] \cos \beta_0 + 2 \sqrt{r_1 r_2} 
\sqrt{(r - r_1)(r_2 - r)} \sin \beta_0 \right\}, \eqno (19)$$
$$\varphi = \varphi_0 + {1 \over 2} {m \over \vert M \vert}
\left\{ \sin^{-1} \left[ {1 \over \cos \theta_0} \left( -1 + 
{\sin^2 \theta_0 \over {1 + \cos \theta}} \right) \right]
- \sin^{-1} \left[ {1 \over \cos \theta_0} \left( -1 + 
{\sin^2 \theta_0 \over {1 - \cos \theta}} \right) \right] 
\right\},$$
where $t_0$, $\varphi_0$ and $\beta_0$ are integration contants (in 
addition to the constants of motion $E$, $K$ and $m$).

\sa

For $Q=0$, we recover the well-known result for the Coulomb system : 
all finite trajectories are planar and closed 
(ellipses) and the motion 
has the Kepler period $T_C = 2 \pi Z (-2E)^{-3/2}$. For 
$Q > 0$, the coordinates $r$ and $z$ have the period $T_r = T_z = 
T_C$ which is generally not commensurable with $T_{\varphi}$. 
Thus for the overall motion to be periodic, rather than only 
quasi-periodic, we again arrive at the commensurability 
condition (12) (an important qualification not mentioned in 
our previous article [13]). The overall period $T$ in this case 
is
$$
T = k_1 T_C. 
\eqno (20)
$$
Therefore, the finite trajectories for $V_C$ are not closed in 
general ; however, all 
trajectories are quasi-periodic and the periodicity evocated 
in ref.~[13] is actually a quasi-periodicity. 

\sb

\noindent{\bf 3. Conclusions}

\sa

In this paper we have concentrated on the properties of 
super-integrable, but not maximally super-integrable, systems 
in classical mechanics. We have used the examples of two ring 
shaped potentials, $V_O$ (eq.~(2)) and $V_C$ (eq.~(3)), 
to show that the classical trajectories are 
always quasi-periodic and that they are periodic if a 
commensurability condition (eq.~(12)) is satisfied. (It is 
remarkable that the same closure condition (12) be obtained for 
$V_O$ and $V_C$.) For both 
potentials, periodicity depends on only one of the three constants 
of the motion, namely the constant $m$, the value of the 
angular momentum component $L_z$. Whether the trajectory lies 
in a plane or not may depend on the values of other integrals of 
motion, a fact to be explored in a publication in preparation 
[18].

\sa

Periodic trajectories exist in Hamiltonian systems that are not 
super-integrable, but they are rare. Here, we have infinitely 
many periodic trajectories. Moreover, since an irrational 
number can be approximated with any chosen accuracy by a 
rational one, infinitely many periodic trajectories lie in the 
neighbourhood of any non-periodic one.

\sa

Turning to the quantum mechanical systems corresponding to the 
potentials $V_O$ and $V_C$, we note that, as usual, 
super-integrability manifests itself in the degeneracy of energy
levels. The fact that these systems are not maximally super-integrable has 
the consequence that the degeneracy is not maximal either. Indeed, the wave 
functions depend on three quantum numbers, say the eigenvalues of the quantum 
versions of $H$, $A_2$ and $A_3$ or $H$, $B_1$ and $B_2$ for 
$V_O$ or $V_C$, respectively. The quantization conditions 
then imply that states with different eigenvalues of $A_2$, 
repectively $B_1$, have the same energy, which does however depend 
(via $M$) on the other quantum number $m$. The Lie algebra ($su(2)$) 
explaining the degeneracy is of rank one in both cases [13,14]. For a subset 
of states satisfying the periodicity condition (12), a higher degeneracy, as 
yet not investigated, will occur.

\sa

Let us mention that new super-integrable Hamiltonian systems 
have recently been found by Evans [19] and that it would be 
of interest to study the corresponding classical and quantum 
solutions.

\sa

We plan to further investigate properties of super-integrable 
Hamiltonian systems, partly in view of the possible application 
of such systems with infinitely many periodic trajectories in 
particle accelerators and storage rings. The classical 
trajectories in the ring shaped potentials $V_O$ and $V_C$ will be 
given in graphical form in ref.~[18].

\sb

\noindent{\bf Acknowledgements}

\sa

The work reported here was finished during M.~K.'s visit to the 
Centre de recherches math\'ematiques, Universit\'e de 
Montr\'eal, made possible by a Coop\'eration Qu\'ebec-France 
research grant (Projet 20~02~20~89). The research of P.~W. is 
partially supported by research grants from NSERC of Canada and 
FCAR du Qu\'ebec. The authors thank their co-author of 
ref.~[18], Dr. G.~H. Lamot from the Institut de Physique 
Nucl\'eaire de Lyon, and Dr. M.~A. Rodriguez from the University 
of Madrid, for helpful discussions. They are indebted to Dr. N.~W. 
Evans from the Queen Mary College (London) for sending them a 
preprint of ref.~[19].

\vfill\eject

\noindent{\bf References}

\sa

\parindent = 0.8 true cm
  \baselineskip = 0.75 true cm

\item{[1]} J. Bertrand, C. R. Acad. Sci. (Paris) 77 (1873) 849.

\item{[2]} S. Wojciechowski, Phys. Lett. A 95 (1983) 279 ; see 
also : Phys. Lett. A 64 (1977) 273.

\item{[3]} W. Pauli, Jr.,  Z. Phys. 36 (1926) 336. 

\item{[4]} V. A. Fock,     Z. Phys. 98 (1935) 145.

\item{[5]} V. Bargmann,    Z. Phys. 99 (1936) 576.

\item{[6]} J.~M. Jauch and E.~L. Hill, Phys. Rev. 57 (1940) 641 ; 
Yu.~N. Demkov, Zh. Eksperim. Teor. Fiz. 26 (1954) 757 ; G.~A. 
Baker, Jr., Phys. Rev. 103 (1956) 1119.

\item{[7]} M. Moshinsky, The harmonic oscillator in modern 
physics : From atoms to quarks (Gordon and Breach, N. Y., 1969).

\item{[8]} P. Winternitz, Ya.~A. Smorodinski\u \i, 
M. Uhl\'\i \v r and 
I. Fri\v s, Yad. Fiz. 4 (1966) 625 [Sov. J. Nucl. Phys. 4 (1967) 444].  

\item{[9]} A.~A. Makarov, J.~A. Smorodinsky, Kh. Valiev and P. 
Winternitz, Nuovo Cimento A 52 (1967) 1061.

\item{[10]} W. Miller, Jr., Symmetry and separation of 
variables (Addison Wesley, Mass., 1977).

\item{[11]} H. Hartmann, Theoret. Chim. Acta (Berl.) 24 (1972) 
201 ; H. Hartmann and D. Schuch, Int. J. Quantum Chem. 
18 (1980) 125.

\item{[12]} M. Kibler and T. N\'egadi, Int. J. Quantum Chem. 26 
(1984) 405 ; Croat. Chem. Acta 57 (1984) 1509.

\item{[13]} M. Kibler and P. Winternitz, J. Phys. A : Math. Gen. 
20 (1987) 4097. 

\item{[14]} C. Quesne, J. Phys. A : Math. Gen. 21 (1988) 3093.

\item{[15]} A. Guha and S. Mukherjee, J. Math. Phys. 28 (1987) 
840 ; M. Kibler and T. N\'egadi, Phys. Lett. A 124 (1987) 42 ; 
I. S\"okmen, Phys. Lett. A 132 (1988) 65 ; 
L. Chetouani, L. Guechi and T.~F. Hammann, J. Math. Phys. 30 (1989) 
655.

\item{[16]} M. Kibler and T. N\'egadi, in : Group theoretical 
methods in physics, eds. Y. Saint-Aubin and L. Vinet (World 
Scientific, Singapore, 1989) p.~666. 

\item{[17]} M. Born, The mechanics of the atom (Bell and Sons, 
London, 1927).

\item{[18]} M. Kibler, G.~H. Lamot and P. Winternitz, to be 
published.

\item{[19]} N.~W. Evans, {\it Super-integrability in classical mechanics}, 
to appear in Phys. Rev. A.

\bye